\documentclass{article}
\usepackage{spconf,amsmath,graphicx}
\usepackage{color}
\usepackage[shortlabels]{enumitem}
\usepackage{amssymb}
\usepackage{makecell}
\usepackage{subfigure}
\usepackage{url}
\usepackage[dvipsnames]{xcolor}

\urldef{\wav2vec}\url{https://github.com/pytorch/fairseq/tree/master/examples/wav2vec}
\urldef{\decoar}\url{http://github.com/awslabs/speech-representations}

\definecolor{mygray}{gray}{0.6}

\title{
Probing Acoustic Representations for Phonetic Properties
}
\name{Danni Ma$^{1}$ ~Neville Ryant$^{2}$ ~Mark Liberman$^{2}$}
\address{$^{1}$Department of Computer and Information Science, University of Pennsylvania\\
$^{2}$Linguistic Data Consortium, University of Pennsylvania}
\begin{document}
%\ninept
%
\maketitle
\begin{abstract}
Pre-trained acoustic representations such as wav2vec and DeCoAR have attained impressive word error rates (WER) for speech recognition benchmarks, particularly when labeled data is limited. But little is known about what phonetic properties these various representations acquire, and how well they encode transferable features of speech. We compare features from two conventional and four pre-trained systems in some simple frame-level phonetic classification tasks, with classifiers trained on features from one version of the TIMIT dataset and tested on features from another.  All contextualized representations offered some level of transferability across domains, and models pre-trained on more audio data give better results; but overall, DeCoAR, the system with the simplest architecture, performs best.   This type of benchmarking analysis can thus uncover relative strengths of various proposed acoustic representations.
\end{abstract}

\section{Introduction}
Inspired by the success of pre-trained word representations \cite{mikolov2013distributed, pennington2014glove}, there has been increasing interest in \textit{unsupervised} learning of distributed vector representations from acoustic data, which allow representations to be pre-trained once and then used repeatedly for other tasks. These models \cite{bengio2014word, kamper2016deep, chung2016audio, he2016multi} aim to map acoustic sequences to a latent embedding space in which audio segments that have similar linguistic properties (phonetic, phonological, lexical, etc) are closer than segments with divergent properties.

More recent work has explored incorporating contextual information in the pre-training stage, and model the use of frames in context of the entire input sequence. The pre-training objectives, usually using self-supervised learning, include next step prediction \cite{oord2018representation, schneider2019wav2vec}, masked acoustic modeling \cite{song2019speech, liu2020mockingjay, chi2020audio}, and connectionist temporal classification \cite{ling2020bertphone}. Pre-trained contextualized acoustic representations appear to be extremely effective. For example, wav2vec 2.0 \cite{baevski2020wav2vec} and DeCoAR \cite{ling2020deep} have attained state-of-the-art results for speech recognition on corpora such as Wall Street Journal (WSJ; \cite{garofalo2007csr}) and LibriSpeech \cite{panayotov2015librispeech}. More impressively, they produce competitive results even when the amount of available labeled data is low -- e.g., wav2vec 2.0 achieves a 4.8\% WER for LibriSpeech {\it test-clean} using only \textit{10 minutes} of labeled data.
% They have yielded great improvements on several tasks, such as speech recognition \cite{baevski2019vq}, speaker verification \cite{ling2020bertphone} and language identification \cite{cai2018exploring}.

% Past work
These gains in automatic speech recognition (ASR) performance show that pre-trained representations encode high-level abstractions of acoustic sequences. However, relatively little is known about the nature of the linguistic information encoded by these abstractions. One notable exception is \cite{pascual2019learning}, which finds that the hidden layers of these networks outperform traditional features for a range of tasks, including speaker identification, emotion classification, and speech-to-text. \cite{li2020does} found that the learned features become progressively more abstract at higher layers, normalizing for dimensions such as speaker, channel, and environmental conditions. In response to this sparse landscape, we asked the following questions: 
\begin{enumerate}[(1)]
    \item At what level of granularity can pre-trained representations capture phonetic knowledge?
    \item Do pre-trained representations encode phonetic properties of speech  better than conventional acoustic features such as MFCCs and Mel filterbanks?
    \item How domain-invariant are pre-trained representations; i.e., to what extent do models trained using these features degrade when faced with a new, unseen domain?
\end{enumerate}

%%%%%%%%%%%%%%%%
%%% Figure 1 %%%
%%%%%%%%%%%%%%%%
\begin{figure}[t]
    \begin{minipage}[b]{1.0\linewidth}
      \centering
      \centerline{\includegraphics[width=\linewidth]{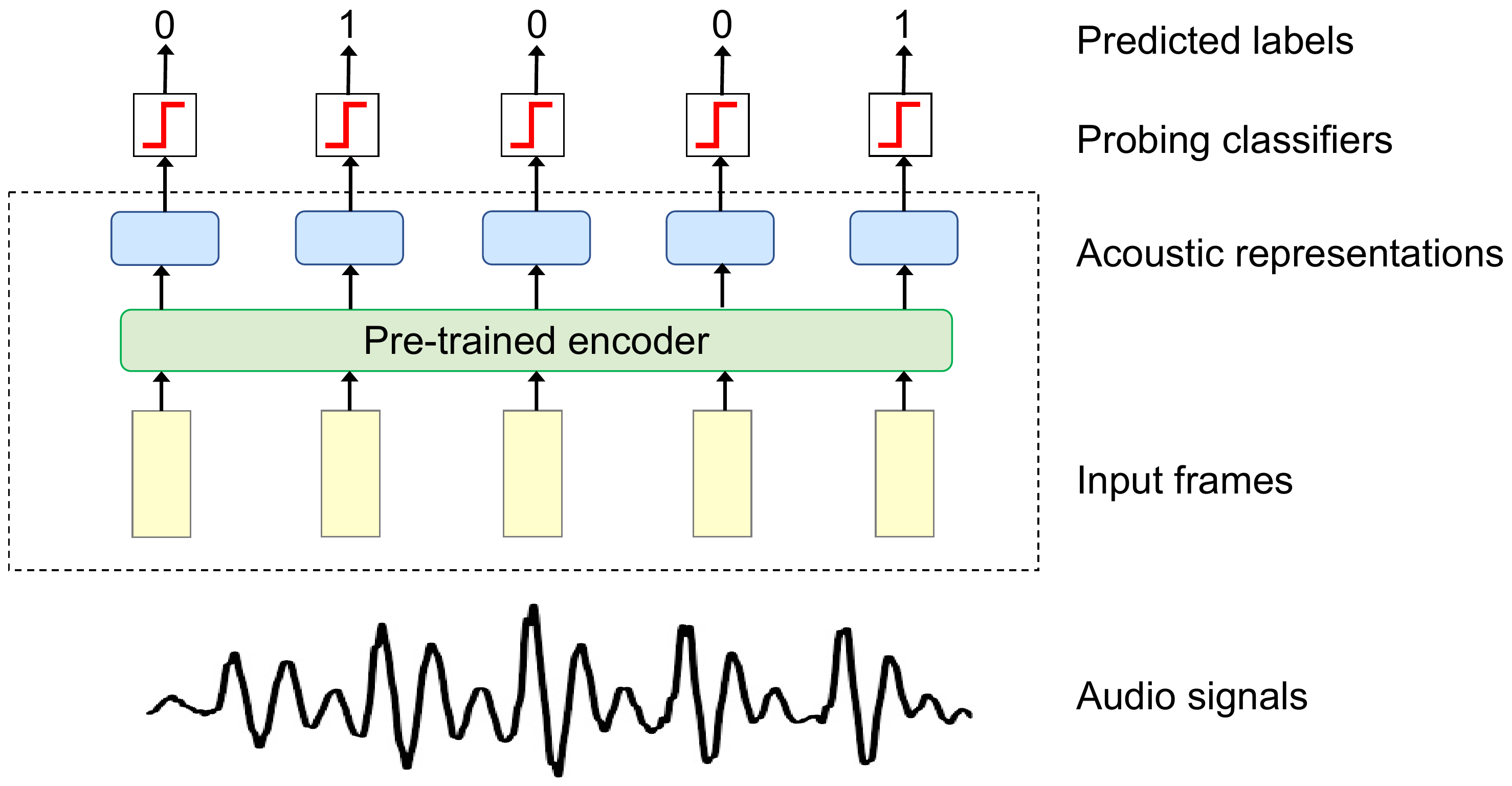}}
    \end{minipage}
    \caption{Probing experiment model architecture.}
    \label{fig:architecture}
\end{figure}

Inspired by \cite{liu2019linguistic, conneau2018you}, we address these questions via a series of probing experiments, which attempt to measure how well information about phonetic structure can be extracted from representations. Each experiment has the same format: a simple classifier attempts to predict frame-wise labels using the last layer of a pre-trained encoder as features. Performance of these classifiers is taken as a proxy for how well the representation encodes the relevant phonetic differences; i.e., if a simple classifier is able to successfully perform phone classification using {\bf only} the pre-trained encoder's output as features, this is evidence that the encoder has learned relevant phonetic properties. For a visual depiction of this architecture, see Figure~\ref{fig:architecture}.

Using this paradigm, we produce a systematic comparison between several popular pre-trained acoustic representations. We analyze both their capacity for encoding phonetic information at different levels of granularity -- speech, vowel, and phone -- as well as their ability to generalize across domains. Our experimental results reveal the following findings:
\begin{enumerate}[(1)]
    \item All pre-trained representations outperform conventional acoustic features for these tasks. 
    \item For all representations, performance on the probing tasks drops as the granularity of the phonetic knowledge required grows finer. For example, classifiers perform best on speech activity detection, and worst for phone classification.
    %
    % On the other hand, probing results vary when using different representations, even if they are pre-trained on the same data with similar objectives.
    \item The different pre-trained representations differ dramatically in how well they perform, despite being conceptually similar and using the same pre-training data.
    \item Pre-trained representations are more domain invariant than conventional acoustic features. Across classification tasks, the drop in performance when there is a mismatch between train/test domain is far lower for pre-trained encoders such as DeCoAR than for conventional acoustic features. 
\end{enumerate}

\label{sec:encoders}
\section{Pre-trained acoustic representations}
We consider four pre-trained acoustic representations\footnote{We focus on these models because they are most likely to be used by researchers, and their pre-trained weights and code are publicly available.}:

\begin{itemize}
    \item \textbf{wav2vec} \cite{schneider2019wav2vec} is an extension of {\it word2vec} \cite{mikolov2013distributed} to the audio domain. It consists of a multi-layer CNN operating on raw speech samples and optimized using noise contrastive estimation. We use {\it fairseq}'s \cite{ott2019fairseq} {\it wav2vec\_large} model. 
    \item \textbf{vq-wav2vec} \cite{baevski2019vq} is an extension of {\it wav2vec} that adds a self-supervised prediction task. In a first step, discrete labels are assigned to each frame by quantizing the dense outputs of a {\it wav2vec} encoder using either a Gumbel-Softmax or k-means clustering. This label sequence is then used as input to BERT \cite{devlin2019bert} pre-training and the hidden activations of the resulting BERT model are used as the acoustic representation. We use the {\it bert\_kmeans} model distributed with {\it fairseq} \cite{ott2019fairseq}.
    \item \textbf{Mockingjay} \cite{liu2020mockingjay} is a direct adaptation of BERT to the acoustic domain. A transformer is trained to reconstruct masked filterbank outputs using an L1 loss function. We use the implementation from the S3PRL toolkit \cite{S3PRL} and the {\it LinearLarge-libri} checkpoint.
    \item \textbf{DeCoAR} \cite{ling2020deep} is inspired by ELMo \cite{peters2018deep}. Like Mockingjay, it is a bidirectional encoder trained under a reconstruction loss, though it uses a bidirectional LSTM instead of a transformer as its encoder. Conceptually, it is the simplest of the pre-trained representations. We use the implementation from Amazon's {\it speech-representations} GitHub repo\footnote{\decoar} with the {\it decoar-encoder-29b8e2ac} checkpoint. 

\end{itemize}
Basic information about these four representations, including output dimensionality and pre-training corpus, are available in Table~\ref{tab:rep}.

In addition, we consider two non-pretrained acoustic representations:

\begin{itemize}
    \item \textbf{MFCC} -- 40-D Mel frequency cepstral coefficients (MFCCs)
    \item \textbf{fbank} -- 40-D Mel scale filterbank outputs
\end{itemize}
The MFCC and filterbank features are extracted using {\it librosa} \cite{mcfee2015librosa} with a 10 ms step size and a 35 ms analysis window. For both feature types, we concatenate an 11-frame context (5-1-5), yielding a final feature dimension of 440.

%%%%%%%%%%%%%%%%%%
% Table
%%%%%%%%%%%%%%%%%%
\begin{table}[htb]
\small
\centering
\scalebox{0.95}{\begin{tabular}{ccccc}
    \Xhline{2\arrayrulewidth}
    \textbf{Model} & \textbf{Dimensionality} & \textbf{Encoder} & \textbf{Unlabeled data} \\ 
    \Xhline{2\arrayrulewidth}
    wav2vec & 512 & CNN & 960H LibriSpeech \\
    \hline
    vq-wav2vec & 768 & CNN & 960H LibriSpeech \\
    \hline
    Mockingjay & 768 & Transformer & 360H LibriSpeech \\
    \hline
    DeCoAR & 2048 & Bi-LSTM & 960H LibriSpeech \\
    \hline
\end{tabular}}
\caption{Embedding size and encoder type for the pre-trained acoustic representations. All models are trained on either 360 hours or 960 hours of LibriSpeech.}
\label{tab:rep}
\end{table}
\label{sec:setup}
\section{Probing Experiments}
% \begin{table*}[t]
% \centering
% \scalebox{1}{\begin{tabular}{c||c|c|c|c|c|c}
% \Xhline{2\arrayrulewidth}
% \textbf{Dataset} & TIMIT & NTIMIT & CTIMIT & FFMTIMIT & STC-TIMIT & WTIMIT \\ 
% \Xhline{2\arrayrulewidth}
% % LDC Catalog No. & LDC93S1 & LDC93S2 & LDC96S30 & LDC96S32 & LDC2008S03 & LDC2010S02 \\
% % \hline
% Sample rate & 16000 & 16000 & 8000 & 16000 & 8000 & 16000 \\
% \hline
% Data source & microphone & telephone & telpephone & microphone & telephone & telephone \\
% \hline
% Bandwidth & wideband & narrowband & narrowband & wideband & narrowband & wideband \\
% \hline
% % Mockingjay & 768 & Transformer & 360 Libri. \\
% % \hline
% % DeCoAR & 2048 & Bi-LSTM & 960h Libri. \\
% % \hline
% \end{tabular}}
% \caption{TIMIT family of corpora used in this paper.}
% \label{tab:datasets}
% \end{table*}

\begin{table*}[htb]
\centering
\scalebox{0.86}{\begin{tabular}{c||c|c|c|c|c|c|c|c|c|c|c|c|c|c|c}
\Xhline{2\arrayrulewidth}
\textbf{Task} & \multicolumn{3}{c|}{\textbf{SAD}} & \multicolumn{3}{c|}{\textbf{vowel detection}} & \multicolumn{3}{c|}{\textbf{sonorant detection}} & \multicolumn{3}{c|}{\textbf{fricative detection}} & \multicolumn{3}{c}{\textbf{phone classification}}\\
\hline
Classifier & LR & SVM & NN & LR & SVM & NN & LR & SVM & NN & LR & SVM & NN & LR & SVM & NN \\
\Xhline{2\arrayrulewidth}
\multicolumn{16}{l}{\textit{Baseline representations}}\\
\hline
Majority & \multicolumn{3}{c|}{92.48} & \multicolumn{3}{c|}{60.92} & \multicolumn{3}{c|}{72.14} & \multicolumn{3}{c|}{28.47} & \multicolumn{3}{c}{1.13} \\
\hline
% \makecell{fbank \\ (no context)} & 89.55 & 85.45 & 91.28 & 81.27 & 73.18 & 83.97 & 90.08 & 89.25 & 91.57 & 57.02 & 52.02 & 66.98 & 29.58 & 7.67 & 31.45 \\
% \hline
fbank & 93.18 & 87.05 & 96.48 & 84.83 & 78.65 & 89.03 & 91.93 & 90.80 & 94.28 & 59.70 & 56.20 & 75.40 & 38.25 & 15.92 & 46.10 \\
\hline
% \makecell{MFCC \\ (no context)} & 89.62 & 92.25 & 91.48 & 81.02 & 65.80 & 84.30 & 89.93 & 88.58 & 91.90 & 56.97 & 49.27 & 66.35 & 29.38 & 12.82 & 33.07 \\
% \hline
MFCC & 93.33 & 85.8 & 96.32 & 84.68 & 77.32 & 88.98 & 91.77 & 88.53 & 94.42 & 60.27 & 50.17 & 74.98 &38.02 & 17.65 & 46.00 \\
\Xhline{2\arrayrulewidth}
\multicolumn{16}{l}{\textit{Pre-trained representations}}\\
\hline
wav2vec & 97.08 & 97.18 & 97.67 & 87.92 & 87.92 & 90.43 & 93.55 & 93.43 & 94.72 & 72.97 & 72.45 & 79.42 & 61.63 & 56.50 & 62.18 \\
% \hline
% \color{mygray} \makecell{(old) \\ vq-wav2vec} & \color{mygray} 93.58 & \color{mygray} 93.62 & \color{mygray} 96.62 & \color{mygray} 72.65 & \color{mygray} 72.95 & \color{mygray} 75.85 & \color{mygray} 81.02 & \color{mygray} 81.48 & \color{mygray} 83.62 & \color{mygray} 45.02 & \color{mygray} 45.03 & \color{mygray} 49.45 & \color{mygray} 13.27 & \color{mygray} 7.97 & \color{mygray} 14.25 \\
\hline
vq-wav2vec & 97.21 & 97.25 & 97.83 & 88.61 & 88.67 & 90.81 & 93.98 & 93.98 & 94.86 & 74.67 & 74.94 & 80.20 & 65.00 & 61.65 & 65.69 \\
\hline
% \color{mygray}
% \makecell{Mockingjay \\ (step = 0.01)} & \color{mygray} 95.03 & \color{mygray} 95.75 & \color{mygray} 96.88 & \color{mygray} 59.13 & \color{mygray} 61.05 & \color{mygray} 63.65 & \color{mygray} 69.02 & \color{mygray} 70.15 & \color{mygray} 73.90 & \color{mygray} 33.25 & \color{mygray} 33.47 & \color{mygray} 37.75 & \color{mygray} 7.32 & \color{mygray} 5.53 & \color{mygray} 10.78 \\
% \hline
Mockingjay & 96.87 & 96.84 & 97.60 & 84.44 & 84.58 & 86.15 & 90.92 & 91.10 & 91.82 & 67.24 & 67.61 & 73.55 & 40.61 & 33.52 & 47.52 \\
\hline
DeCoAR & 97.72 & 97.63 & \textbf{98.22} & 89.15 & 89.17 & \textbf{91.03} & 94.35 & 94.32 & \textbf{95.18} & 77.53 & 77.62 & \textbf{82.02} & 67.10 & 63.23 & \textbf{67.23} \\
\Xhline{2\arrayrulewidth}
\end{tabular}}
\caption{Average in-domain performance for all probing tasks. Numbers reported are the average of F1 scores on six TIMIT datasets. The best result for each task is bolded. \textit{LR}: logistic regression; \textit{SVM}: max-margin classifier; \textit{NN}: neural network.}
\label{tab:performance}
\end{table*}

\subsection{The prediction tasks}
For our probing tasks, we select five frame-level prediction tasks: speech activity detection (SAD), sonorant detection, vowel detection, fricative detection and phone classification. The first four tasks are binary classification tasks, which require determining whether or not a frame belongs to a chosen phonetic class (e.g., speech vs. non-speech or sonorant vs. non-sonorant). The last is a multiway classification task that requires determining which of 39 phones a frame belongs to. Together, these tasks cover a range of phonetic phenomena differing greatly in their granularity, ranging from the superficial (distinguishing speech from non-speech) to very fine-grained (distinguishing between individual phones).

Frame labels are assigned using the manual phone-level segmentation distributed with TIMIT. For the binary classification tasks, the target classes are defined as follows:
    \begin{itemize}
        \item \textbf{fricative}: ch, dh, f, hh, jh, s, sh, th, v, z, zh
        \item \textbf{vowel}: aa, ae, ah, ao, aw, ax, ax-h, axr, ay, eh, el, em, en, eng, er, ey, ih, ix, iy, ow, oy, uh, uw, ux
        \item \textbf{sonorant}: aa, ae, ah, ao, aw, ax, ax-h, axr, ay, eh, el, em, en, eng, er, ey, ih, ix, iy, l, m, n, ng, nx, ow, oy, r, uh, uw, ux, w, y
        \item \textbf{speech}: aa, ae, ah, ao, aw, ax, ax-h, axr, ay, b, bcl, ch, d, dcl, dh, dx, eh, el, em, en, eng, er, ey, f, g, gcl, hh, hv, ih, ix, iy, jh, k, kcl, l, m, n, ng, nx, ow, oy, p, pcl, q, r, s, sh, t, tcl, th, uh, uw, ux, v, w, y, z, zh
    \end{itemize}
For the phone classification task, we train using the full 61 phone set, then map to the standard 39 phone set used for TIMIT phone classification experiments \cite{lee1989speaker}.

\subsection{Datasets}
\label{sec:dataset}

\begin{figure*}[t]
		\centering
		  \subfigure[In-domain and cross-domain performance of representations]{
			\centering	\includegraphics[width=0.55\linewidth]{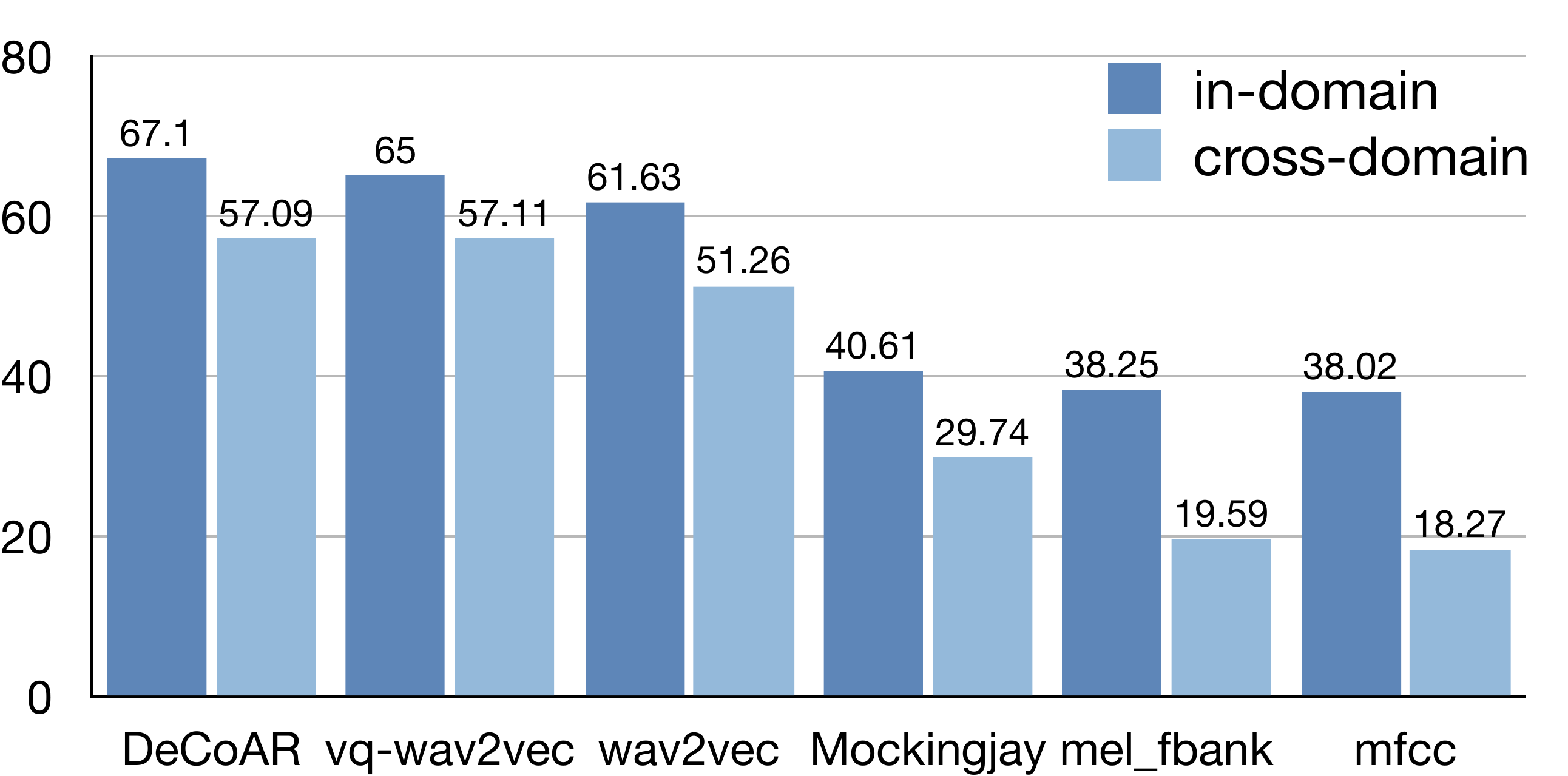}
			\label{fig:representation}}
     	\subfigure[Cross-domain performance for each pair of datasets]{
			\centering	\includegraphics[width=0.41\linewidth]{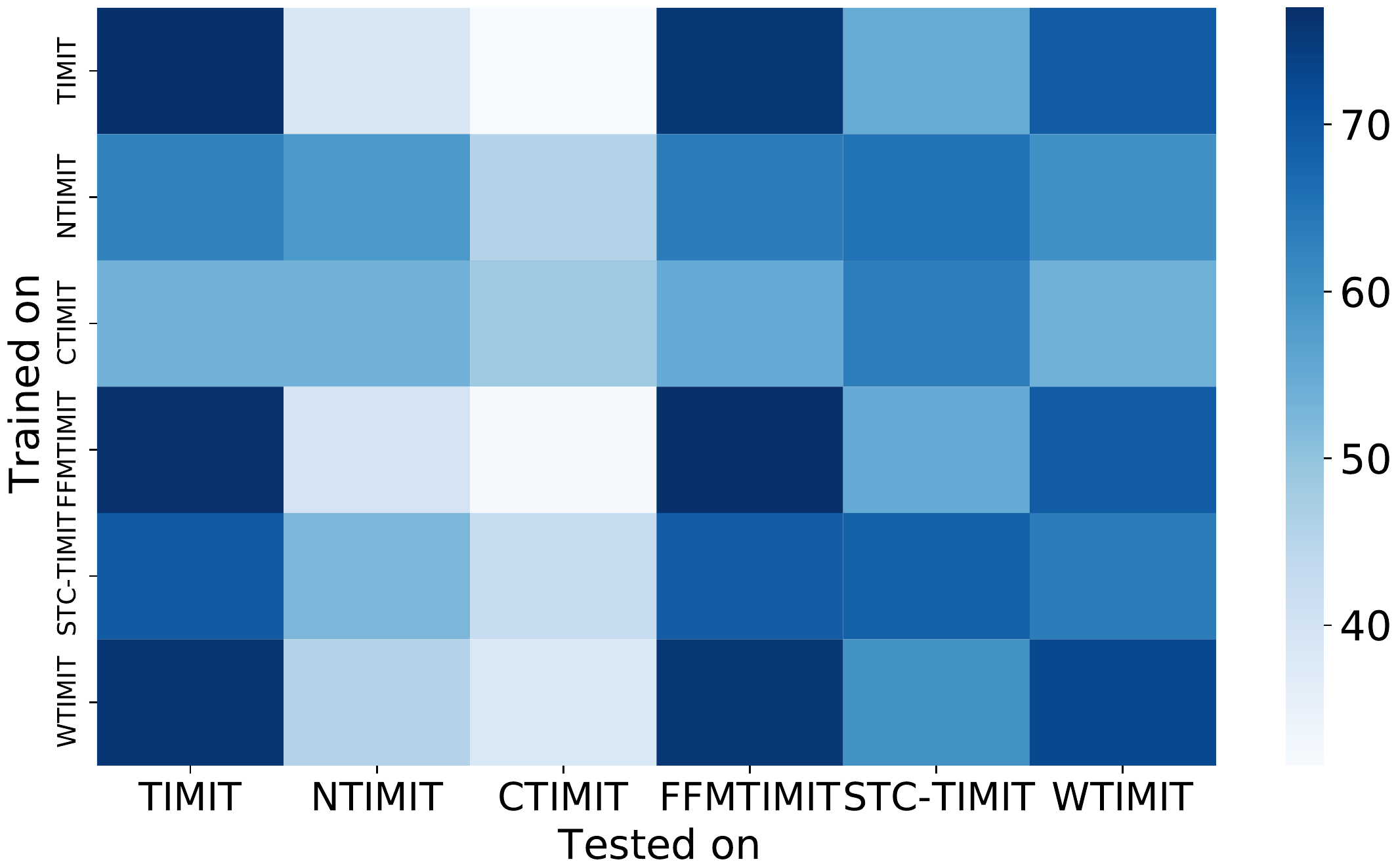}
			\label{fig:cross-domain}}
\caption{Macro-averaged F1 scores for \textbf{phone classification} task. {\it Left:} The left bar in each subgroup represents the average in-domain performance (i.e., the training and test set are from the same dataset). The right bar represents the average cross-domain performance. {\it Right:} Cross-domain performance of DeCoAR. Each cell represents one train/test set combination. Darker colors indicate higher F1.}
\label{fig:avg_performance}
\end{figure*}

For our probing experiments, we utilize the standard TIMIT \cite{garofolo1993timit} plus five TIMIT derivatives:

\begin{itemize}
    \item \textbf{NTIMIT} \cite{fisher1993ntimit}  --  derived by retransmitting the original TIMIT utterances over a telephone handset and the NYNEX telephone network; each utterance was transmitted on a separate call, so there is large variation in channel conditions
    \item \textbf{CTIMIT} \cite{george1996ctimit}  --  generated by transmitting TIMIT over cellular telephone handsets; the transmitting handset was located inside an acoustically isolated cage mounted inside a van driving around New England and the corpus exhibits many transmission related artifacts such as crosstalk, dropout, and low signal-to-noise ratio (SNR)
    \item \textbf{FFMTIMIT} \cite{garofolo1996ffmtimit}  --  alternate free-field microphone recordings from the original TIMIT recording sessions
    \item \textbf{STC-TIMIT} \cite{morales2008stctimit}  --  similar to NTIMIT, but all recordings sent through the same telephone channel
    \item \textbf{WTIMIT} \cite{bauer2010wtimit}  --  retransmission of the TIMIT files over a 3G adaptive multi-rate wideband (AMR-WB) mobile network using Nokia 6220 handsets; much higher quality than CTIMIT
\end{itemize}
NTIMIT and STC-TIMIT are narrowband speech, while the remaining variants are wideband. All experimental results are reported using the full test set.

\subsection{Probing classifiers}
We consider three simple probing classifiers:
\begin{itemize}
    \item \textbf{LR}  --  logistic regression as implemented by {\it sklearn}'s \cite{scikit-learn} {\it LogisticRegression} class
    \item \textbf{SVM}  --  a max-margin classifier fit using {\it sklearn}'s {\it SGDClassifier} class and a hinge loss
    \item \textbf{NN}  --  a simple feedforward neural network consisting of two fully-connected layers of 128 ReLUs. The network was trained for 50 epochs with early stopping using {\it skorch} \cite{skorch}.
\end{itemize}
Because the input representations vary greatly in their dimensionality (ranging from 440 to 2,048), the input features are reduced to 400 dimensions prior to fitting to eliminate this potential confound. Dimensionality reduction is performed by applying singular value decomposition and selecting the top 400 singular components.

For all tasks, we also report the result of a baseline (\textbf{Majority}) that assigns to each frame the most frequent label in the training set.
\section{Experimental Results}
\subsection{Comparison of representations}
\label{sec:rep}
Table \ref{tab:performance} compares the performance of the pre-trained representations and baseline representations across all combinations of probing task and classifier. For the four binary classification tasks, it reports average F1 across all six TIMIT variants. For phone classification it reports the average of macro-averaged (across phone classes) F1 scores. Across the board, the pre-trained representations outperform the baseline representations. For some tasks (SAD, vowel detection, and sonorant detection) the improvement for the pre-trained representations is relatively modest  --  1-5\% absolute. However, for other tasks this advantage is massive with an average improvement of F1 on the order of 15-20\% for fricative detection and 30\% for phone classification. DeCoAR is the strongest performer for all tasks, followed closely by vq-wav2vec and wav2vec, with Mockingjay in fourth place. 

Surprisingly, for two tasks  --  fricative detection and phone classification  --  Mockingjay lags far behind the other pre-trained representations. This deficit is most noticeable for phone classification, where Mockingjay's macro-averaged F1 score is fully 20\% lower (absolute) than the other pre-trained representations and just barely better the conventional representations that serve as a baseline. It is not clear why Mockingjay underperforms DeCoAR, wav2vec, and vq-wav2vec for these two tasks, though we suspect this reflects the fact that Mockingjay was trained only on the {\it train-360} subset of LibriSpeech, which consists entirely of clean data. In contrast, all other models were trained on the {\it train-960} subset, which incorporates another 100 hours of clean data and 500 hours of noisy data. Because Mockingjay was trained on roughly one third as much data, all of which was clean, it may be particularly ill adapted to dealing with noisy and/or bandwidth constrained speech, causing degraded performance for the TIMIT variants, particularly for the harder probing tasks (e.g., phone classification) and TIMIT variants (e.g., CTIMIT).

While the highest F1 scores are universally achieved using the neural network classifier, the same trends are also observed with logistic regression and SVM. Thus to simplify exposition, we present only results from logistic regression in the remainder of this paper.

%%%%%%%%%%%%%%%%%%%%%%%%%%%%%%%%%%
%%% Table: Conditional entropy %%%
%%%%%%%%%%%%%%%%%%%%%%%%%%%%%%%%%%
% \begin{table}[htb]
% \centering
% \scalebox{1}{\begin{tabular}{c|c|c|c|c}
% \Xhline{2\arrayrulewidth}
% \textbf{Task} & \textbf{fricative} & \textbf{vowel} & \textbf{sonorant} & \textbf{SAD} \\ 
% \Xhline{2\arrayrulewidth}
% % \color{mygray} \makecell{Conditional \\ entropy} & \color{mygray} 0.9449 & \color{mygray} 0.8615 & \color{mygray} 0.8350 & \color{mygray} 0.5835 \\
% % \hline
% \makecell{Conditional \\ entropy} & 0.8222 & 0.6665 & 0.6399 & 0.4606 \\
% \hline

% \end{tabular}}
% \caption{In-domain conditional entropy (CE) in bits for binary classification tasks on CTIMIT using Mockingjay. Maximum possible CE: 1.}
% \label{tab:conditional_entropy}
% \end{table}

\subsection{Comparison of probing tasks}
%%%%%%%%%%%%%%%%%%%%%%%%%%%%%%%%
%%% Figure: Confusion Matrix %%%
%%%%%%%%%%%%%%%%%%%%%%%%%%%%%%%%
\label{sec:task}
\begin{figure*}[t]
		\centering
		  \subfigure[Greatest mismatch]{
			\centering	\includegraphics[width=0.48\linewidth]{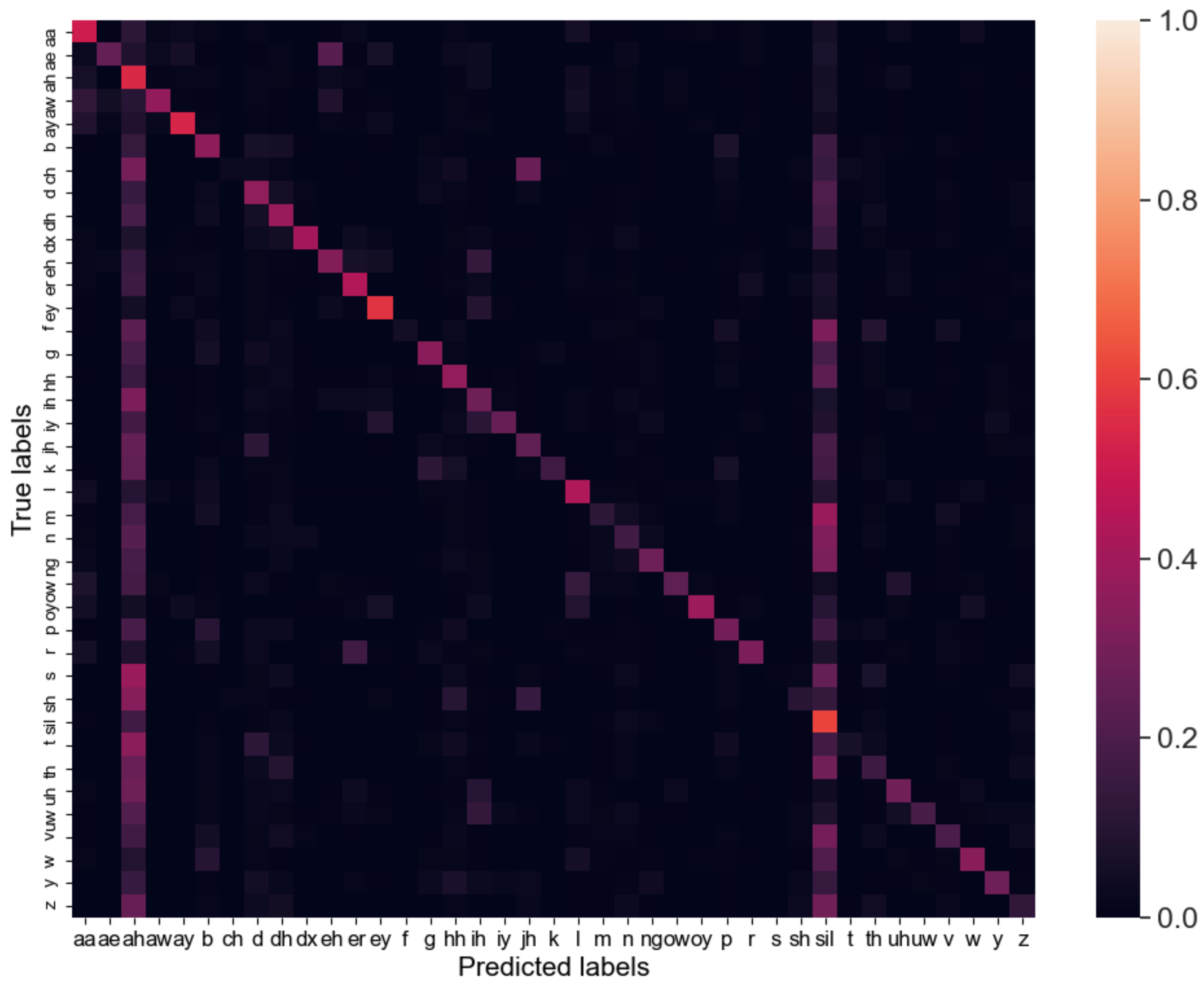}
			\label{fig:greatest}}
     	\subfigure[Least mismatch]{
			\centering	\includegraphics[width=0.48\linewidth]{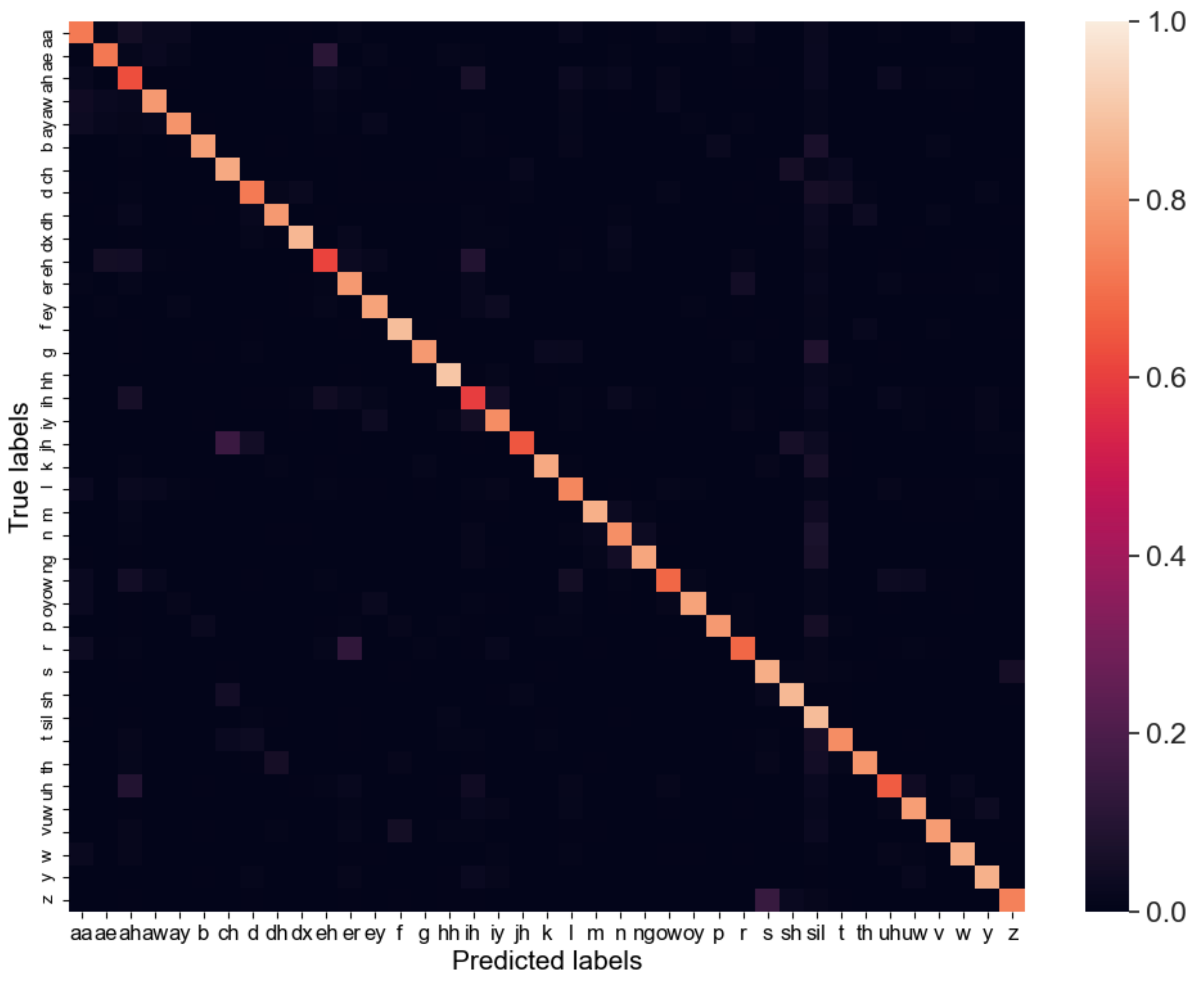}
			\label{fig:least}}
\caption{Confusion matrices for phone classification when using DeCoAR for two extremes of train/test mismatch. {\it Left:} Greatest  mismatch: FFMTIMIT/CTIMIT (CE: \textbf{3.1845}. Maximum possible CE: 5.2854). {\it Right:} Least mismatch: FFMTIMIT/TIMIT (CE: \textbf{1.5975}).}
% Conditional entropy: 1.5975.
\label{fig:mismatch}
\end{figure*}

The probing tasks were designed to require knowledge of phonetic distinctions at different levels of granularity, ranging from the trivial (speech vs. non-speech) to complex (distinguishing between 39 phones). Our prediction was that as the decisions required became more and more granular, difficulty would increase, exposing differences between the representations. As evidenced by Table~\ref{tab:performance}, this is borne out for the probing tasks, with F1 decreasing for all representations as the information required for the task becomes increasingly granular. As expected, SAD is by far the easiest task, with all representations easily exceeding 90\% F1. Vowel detection and sonorant detection are only marginally more taxing, with all representations exceeding 84\% F1. However, the picture becomes somewhat more complex for the last two tasks. For fricative detection, the baseline representations hover around 60\% F1, while F1 for pre-trained representations ranges from 67.24\% for Mockingjay to 77.53\% for DeCoAR. Phone classification is by far the hardest task, with F1 ranging from only 38\% for the baseline representations to a high of 67.1\% for DeCoAR.

\subsection{Domain mismatch}
Up to now we have focused on in-domain performance; that is, how well the representations perform when the probing classifier is trained and tested on the same domain. We now expand our analysis to consider the more common situation where there is a (possibly substantial) mismatch between train and test domains. As phone classification was previously established as the hardest task, we focus on this task as we expect it to be most useful for teasing out differences among the representations.

Figure~\ref{fig:representation} compares average in-domain and cross-domain macro averaged F1 scores for all representations. As expected, cross-domain performance lags in-domain performance across the board with this effect most pronounced for the baseline representations, for which cross-domain F1 is only half of in-domain F1. Generalization to new domains is far better for the pre-trained representations, with the in-domain/cross-domain F1 drop off ranging from 26.8\% (relative) for Mockingjay to only 12-14\% (relative) for vq-wav2vec and DeCoAR. However, even for the pre-trained representations, extreme domain mismatch has a deleterious effect on performance. This is clearly illustrated by Figure~\ref{fig:cross-domain} which depicts F1 for each combination of train/test domain when using DeCoAR. Indeed, training on clean wideband speech (TIMIT) and testing on noisy narrowband, cellular speech (CTIMIT) results in sub 30\% F1, as compared to greater than 70\% F1 when training/testing on clean wideband speech.

% Come back and revise discussion of CE.
To achieve a better understanding of what kinds of errors the probing classifier is making, we ranked the 36 possible train/test domain pairs  by the conditional entropy of resulting classifiers' confusion matrices, then compared the confusion matrices for DeCoAR for the most (train: FFMTIMIT; test: CTIMIT) and least mismatched (train: FFMTIMIT; test: TIMIT) pairs. In the low mismatch condition (Figure~\ref{fig:least}), the overall number of errors is small with most involving pairs of similar fricatives (s/z, ch/jh) or vowels (ae/eh, ih/eh). By contrast, for the high mismatch condition (Figure~\ref{fig:greatest}) the pattern of errors is much more diffuse. While we continue to see errors involving phonetically similar fricatives and vowels, the phones ``sil'' and ``ah'' show a high degree of confusion with almost every other phone. For fricatives, affricates, and stops this is perhaps unsurprising given that the classifier was trained on wideband speech, but tested on narrowband speech. However, it is less clear why so many vowels, nasals, and glides are being misclassified as these two classes.

\section{Conclusion}
% We compare the performance of various acoustic representations on various phonetic classification tasks. These tasks are of different difficulty, and require different granularity levels of phonetic information. We find that probing tasks requiring finer-grained phonetic knowledge are more challenging, and that pre-training enhances generalization ability and cross-domain performance. In addition, we observe a significant performance drop when testing in a noisy target domain, indicating that this is still a major challenge.

% We hope that our analysis will motivate more research on the interpretability of acoustic representations. There are many fascinating directions for future work. First, it is interesting that the system with the simplest architecture, DeCoAR, performs best overall. Broader probes of encoder architecture are therefore warranted. Second, it is worth investigating how pre-training methods affect the generalization ability of representations. Lastly, we hope to see improvement on robustness in new pre-trained representations.

Recently, pre-trained acoustic representations such as DeCoAR \cite{ling2020deep} and wav2vec 2.0 \cite{baevski2020wav2vec} have been leveraged to achieve state-of-the-art on a number of speech-to-text benchmarks. However, little work has been devoted to understanding what types of information these representations actually encode. In this paper, we compare several such representations, as well as two conventional acoustic representations, using a series of phonetic probing tasks spanning multiple levels of phonetic granularity. Our results demonstrate that pre-trained representations capture a wide range of phonetic information, strongly outperforming conventional representations such as MFCCs for all probing tasks. Moreover, classifiers trained using pre-trained representations exhibit marked robustness to mismatch between train and test domain. The system with the simplest architecture, DeCoAR, performed best across all tasks.

In the future, we would like to extend these investigations to encompass additional models (e.g., wav2vec 2.0 and DeCoAR 2.0 \cite{ling2020decoar}) and probes. While this paper was under review, \cite{shah2021all} demonstrated that pre-trained representations may encode a surprising amount of higher level linguistic structure, including aspects of lexical semantics and syntax, suggesting that our probes should be diversified beyond phonetic/prosodic structure.  We also plan to investigate the question of domain generalization for tasks such as speech activity detection and diarization on more realistic real world data such as the CHiME-6 \cite{watanabe2020chime} and DIHARD III \cite{ryant2020third} corpora.

%%%%%%%%%%%%%%%%%%%%%%%%%%%%
% References
%%%%%%%%%%%%%%%%%%%%%%%%%%%%
\nocite{ling2020decoar}
\nocite{shah2021all}
\nocite{ryant2020third}

%\newpage
%\bibliographystyle{IEEEbib}
%\newcommand{\BIBdecl}{\setlength{\itemsep}{0.25 em}}

\bibliographystyle{IEEEtran}

\bibliography{strings,refs}

%%%%%%%%%%%%%%%%%%%%%%%%%%%%
% Additional figures
%%%%%%%%%%%%%%%%%%%%%%%%%%%%
%\clearpage
%\newpage
% \input{09-candidate_figures}

\end{document}